\documentclass[pra,twocolumn,groupedaddress,showkeys]{revtex4-1}
\usepackage{graphicx}
\usepackage{amsmath}
\usepackage{amssymb}
\usepackage[unicode=true,pdfusetitle,
 bookmarks=true,bookmarksnumbered=false,bookmarksopen=false,
 breaklinks=false,pdfborder={0 0 0},backref=false,colorlinks=true,citecolor=blue,urlcolor=violet 
]{hyperref}
\usepackage{xcolor}
\usepackage{caption,subcaption}
\usepackage{physics}
\usepackage{tikz,lipsum,lmodern}
\usepackage[export]{adjustbox}
\usepackage{soul}
\begin{document}
\newcommand{\beq}{\begin{equation}}
\newcommand{\eeq}{\end{equation}}
\newcommand{\orcid}[1]{\href{https://orcid.org/#1}{\resizebox{10px}{!}{\includegraphics{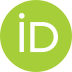}}}}
\bibliographystyle{apsrev}

\title{Understanding Modified Two-Slit Experiments using Path Markers}

\author{Tabish Qureshi\orcid{0000-0002-8452-1078}}
\email{tabish@ctp-jamia.res.in}
\affiliation{Centre for Theoretical Physics, Jamia Millia Islamia, New Delhi-110025, India.}


\begin{abstract}

Some modified two-slit interference experiments were carried out showing
an apparent paradox in wave-particle duality. In a typical such experiment, the
screen, where the interference pattern is supposed to be formed, is replaced
by a converging lens. The converging lens forms the images of the two slits
at two spatially separated detectors. It was claimed that each of these
two detectors give information about which slit a photon came from, even
though they come from the region of interference. These experiments generated
a lot of debate. The various refutations pointed out that the controversial
claims involved some questionable assumptions. However the refutations were
largely philosophical in nature, and one may like to substantiate those with
arguments which are testable, at least in principle.
Here such an experiment is theoretically analyzed
by introducing path markers which are two orthogonal polarization states
of the photon. Analyzing the polarization at the two detectors shows that 
the photons which give rise to interference, and reach a particular detector,
always come from both the slits. This provides clarity in understanding such
experiments by making use of testable quantum correlations.
 
\end{abstract}


\maketitle

\section{introduction}

The two-slit experiment with massive particles, or single photons is probably
the simplest experiment which captures the most intriguing features of
quantum mechanics, especially in a situation where one wants to probe
which of the two slits the particle passed through. 
If the particles show interference, it is not possible to know which slit
each of them passed through. The moment one acquires the knowledge about 
which slit the particle passed through, the interference is lost.
Notable is the fact that no real experimenter or classical apparatus is
needed here - even if the path information gets encoded in another quantum
system, by way of entanglement, that is enough to destroy interference.
Niels Bohr elevated this concept to the status of a separate principle,
the principle of complementarity \cite{bohr}. Feynman believed that this
experiment captures the only mystery in quantum mechanics. It is obvious
that quantum superposition plays an important role in the experiment.
The new understanding that entanglement is at the heart of the 
principle of complementarity \cite{tqrv,triality1,triality2}, reinforces
Feynman's belief.

\begin{figure}
\resizebox{8.0cm}{!}{\includegraphics{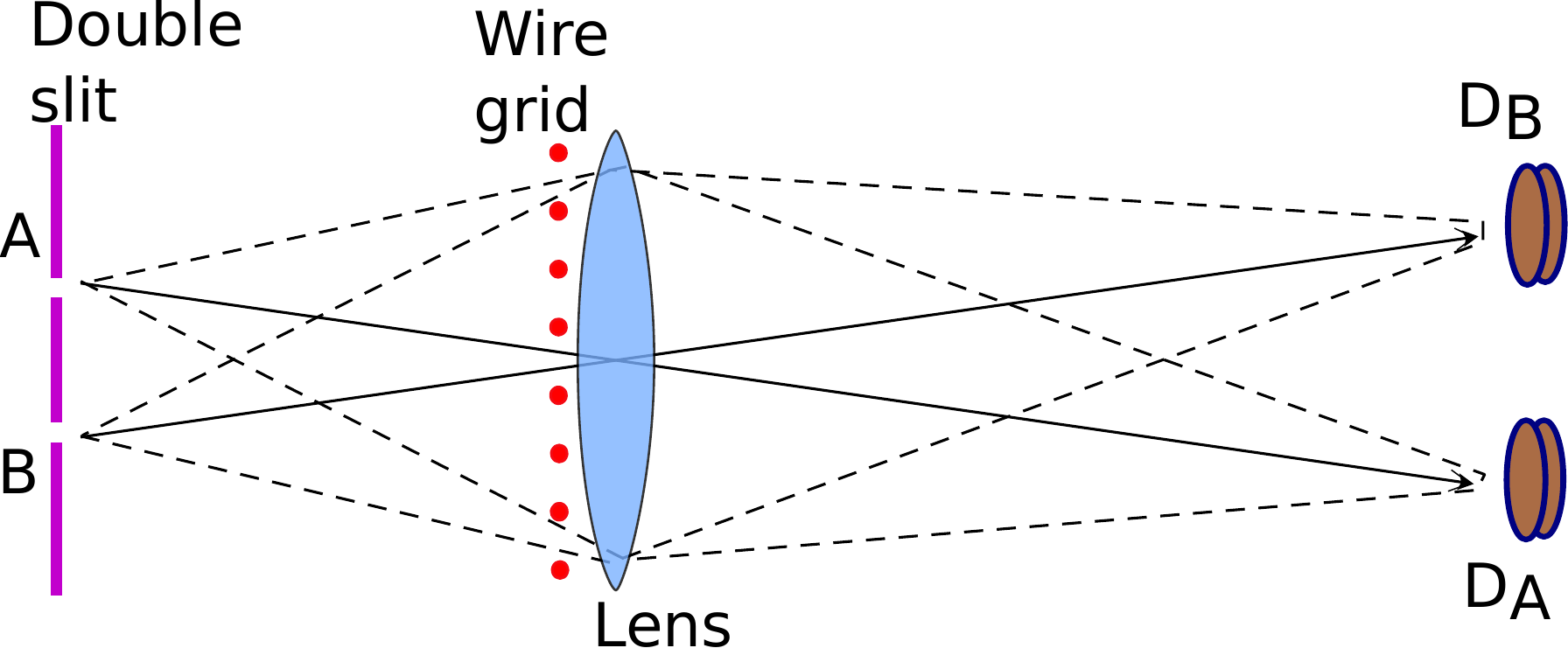}}
\caption{ A schematic representation of a typical modified two-slit experiment.
Photons emerge from two pin-holes (or slits) A and B and interfere. Thin wires
are placed carefully in the exact locations of the dark fringes of the
interference pattern. The lens collects the photon and obtains the
images of the two slits at the two detectors $D_A$ and $D_B$, respectively.}
\label{afsharexpt}
\end{figure}

The principle of complementarity, or wave-particle duality, as it is more
commonly referred to, has stood its ground in the face of various theoretical
and experimental investigations. Some experiments were carried out early
this century which appeared to show a violation this principle
\cite{afshar1, afshar2}, and attracted lot of popular attention \cite{chown}.
A schematic diagram of a typical such experiment is shown in Fig. \ref{afsharexpt}.
It consists of a standard two-slit experiment, with a converging lens just
behind the location where one would put a conventional screen for obtaining
the interference pattern. The experiment uses pinholes instead of conventional
slits, but that does not make any difference. The single photons passing
through the two holes $A$ and $B$, form a sharp interference pattern on
the screen.  If the screen is removed, the light passes through the lens
and produces two images of the holes, which are captured on two detectors
$D_A$ and $D_B$ respectively. Opening only the hole $A$ results in only
detector $D_A$ clicking, and opening only hole $B$ leads to only $D_B$
clicking. The authors argue that the detectors $D_A$ and $D_B$ yield
information about which hole, $A$ or $B$, the photon initially passed through.

Now without a screen one cannot know if the photons interfere and would yield
an interference pattern. The authors devise a clever scheme for
establishing the existence of the interference pattern without actually
observing it. The exact location of the dark fringes is found out by
observing the interference pattern on a screen. Then the screen is removed
and thin wires are placed in the exact locations of the dark fringes.
Their argument was that if the interference pattern exists, sliding in wires
through the dark fringes will not affect the \emph{intensity} of light on
the two detectors. If the interference pattern is not there, the wires would
surely scatter some photons, thus diminishing the photon count at the two
detectors. In this manner, \textcolor{red}{one} can establish the existence of the interference
pattern, without actually disturbing the photons in any way.
The reader would notice the similarity of this scheme with 
the ``interaction-free measurements" where the non-observation of
a particle along one path establishes that it followed the other possible
path, without actually measuring it \cite{kwait}.
The results of the experiments can be summarized as follows.
\begin{enumerate}
\item If only hole $A$ ($B$) is opened, only the detector $D_A$ ($D_B$)
detects photons.
\item If wires are introduced when only one hole is open, the intensity
at the single detector, which received the photons, is reduced.
\item If both holes are opened, both the detector $D_A$ and $D_B$
detect photons.
\item If wires are introduced when both the holes are open, the intensity
at the two detectors remains unaffected, for all practical purposes.
\end{enumerate}
The experiment is quite simple and straightforward. The authors argue that
even when both the holes are open, the detectors $D_A$ and $D_B$
continue to tell us which hole each photon came from. Since the introduction
of the wires does not affect the intensity, it is natural to conclude that
there are dark fringes, and hence an interference pattern, in the region
of spatial overlap of the two photon amplitudes. This leads one to an
apparent paradox in wave-particle duality: the existence of interference should
prohibit any knowledge of which hole a photon passed through, yet the two
detectors at the end appear to provide the same information for every photon.

\begin{figure}
\resizebox{8.0cm}{!}{\includegraphics{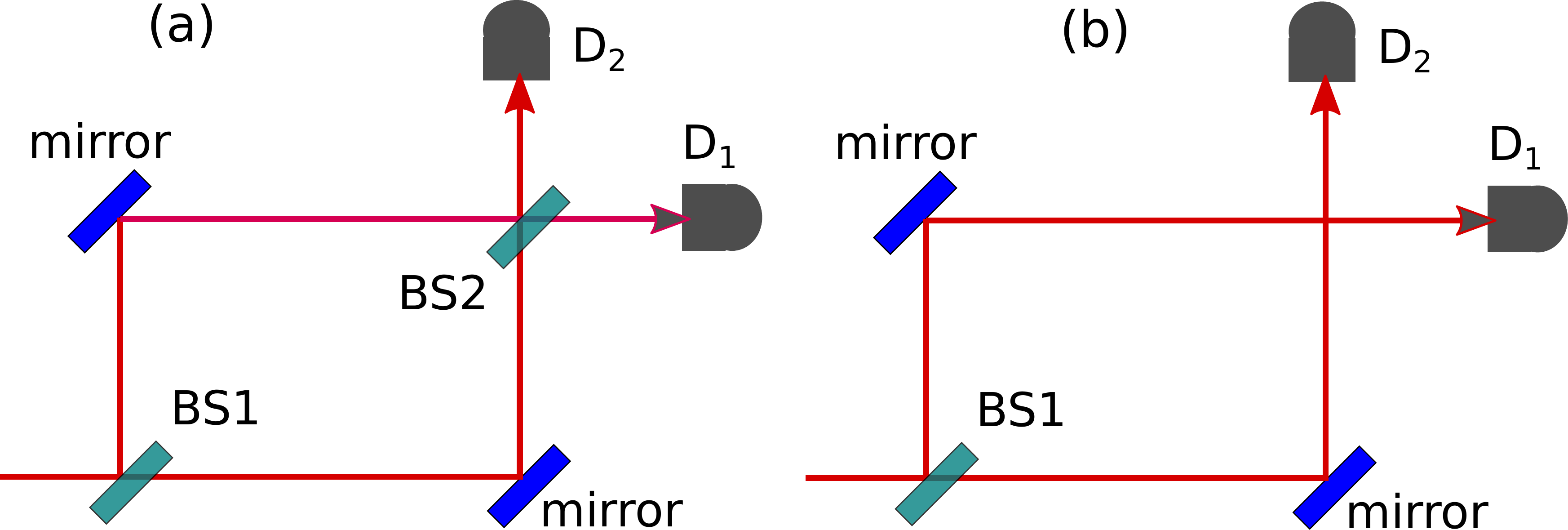}}
\caption{In a Mach-Zehnder interferometer, it is widely believed that 
if the second beam-splitter $BS2$ is removed, the detectors $D_1$ and
$D_2$ give information on which of the two paths a photon followed.}
\label{mz}
\end{figure}

As expected the experiment started a heated debate, with people trying to
find flaws in the experiment \cite{kastner,srinivasan,drezet,steuernagel,
georgiev,flores,jacques,tq,tata,kaloyerou,knight,batelaan}.
However, the various criticisms do not agree among themselves regarding
the perceived flaw in the experiment. The criticism which is in the right
direction is that although the two detectors give which-way information
when only one hole is open, they do not give which-way information when
both the holes are open \cite{tq,kaloyerou,knight}. However, some of
the arguments use the assertion that since the state of the particle
is a superposition of the two paths, there is no which-way information 
to start with \cite{kaloyerou,knight}. While this assertion may be correct
in principle, the assumption
that which-way information is carried by the particle in a Mach-Zehnder
interferometer is widely used today. For instance, in the setup shown in
Fig. \ref{mz}, it is widely believed that if the second beam-splitter is
removed (Fig. \ref{mz}(b)) the detectors $D_1$ and $D_2$ tell us which
path the particle followed, but if BS2 is present (Fig. \ref{mz}(a)),
they do not. Although this belief cannot be defended using standard quantum
mechanics, in most experiments its fallacy does not become apparent.

So the situation is that the controversial claims of these modified two-slit
experiments have been refuted, but mostly on philosophical grounds. The
refuting argument is that it is incorrect to assume that when both the
holes are open, the two detectors give which-way information about
every photon. However, there is no way to prove that the two detectors do
not give which-way information, and a student is bound to be unconvinced
because it is common to accept the same very assumption while dealing with
Mach-Zehnder interferometer (see Fig. \ref{mz}). It is also unavoidable
to have the feeling, mostly stemming from classical prejudice, how can
the detector (say) $D_A$, in Fig. \ref{afsharexpt}, receive photons from
hole $B$?

In the following analysis we introduce quantum path
markers in the two paths of the photons to unambiguously tell which
path the photon followed. If a path marker can distinguish whether a photon
emerged from hole A or B (i.e., by being detected at $D_A$ or
$D_B$), we would like to show that in such a situation, interference cannot
be detected at the lens plane. In addition, we wish to demonstrate that
to get interference at the lens plane, distinguishability by the path
markers must be erased. This is because
a photon that can self-interfere at the lens plane necessarily has wave
components from both holes and may therefore be detected by either $D_A$ or
$D_B$. Using this strategy we analyze these modified
two-slit experiments, and show that the photons which show interference
(via introduction of wires)
and reach a particular detector, always follow \emph{both} the paths.

\section{Two-slit experiment with path markers}

\begin{figure}
\resizebox{8.0cm}{!}{\includegraphics{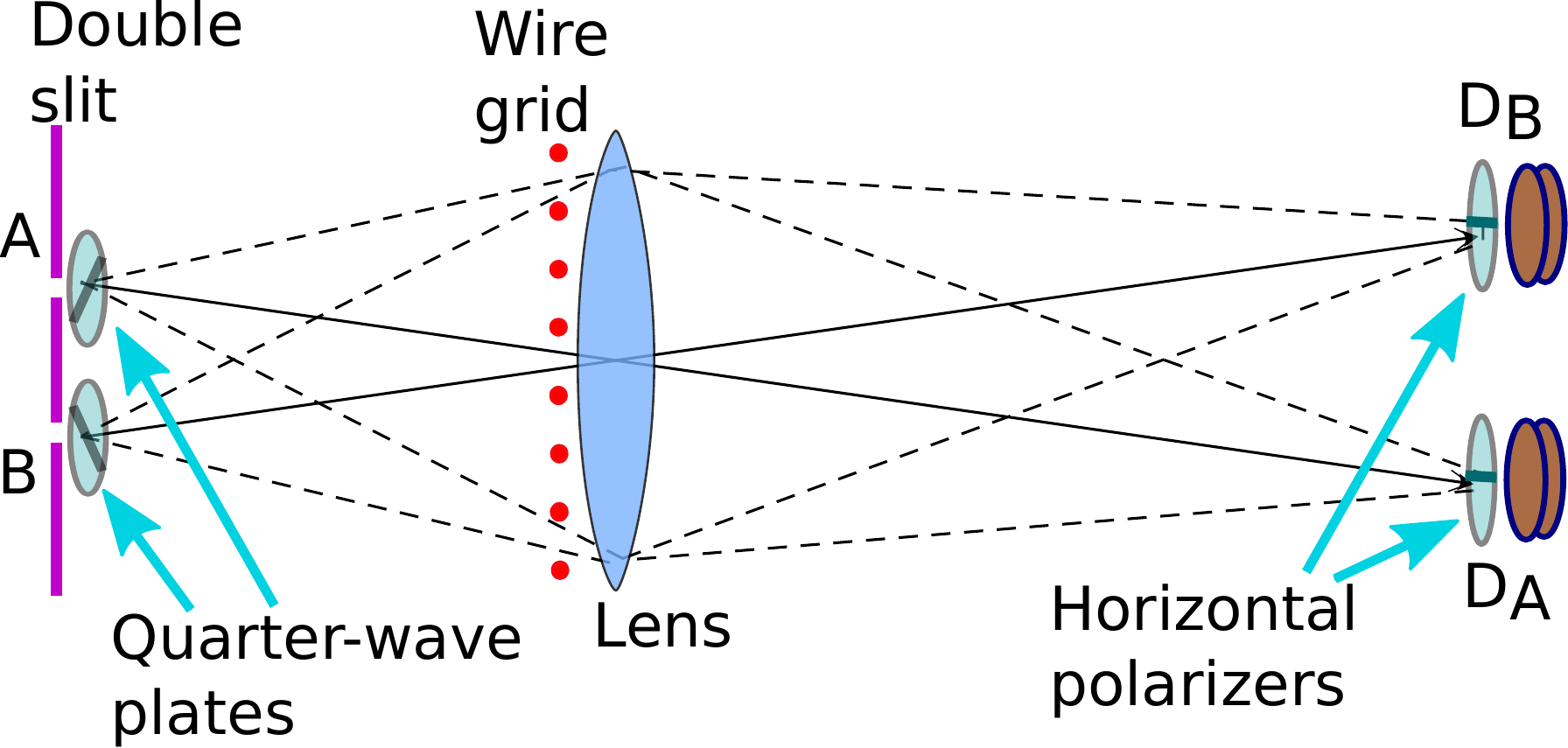}}
\caption{An experiment which is a modification of the one shown in
Fig. \ref{afsharexpt}, by introducing quarter-wave plates in front of the
two slits/holes, and putting horizontal polarizers in front of the
two detectors.}
\label{modafshar}
\end{figure}

Let us assume there is a quantum path-detector which interacts with the
photons as it passes through the two holes. The two orthogonal states of
the path-detector get correlated with the two paths of photons. In the
experiment shown in Fig. \ref{afsharexpt}, this can be achieved by having
a photon source which produces linearly polarized photons, and then 
putting behind the two holes, two quarter-wave plates, which convert the
passing linearly polarized photons to left-circular and right-circular
polarization, respectively (see Fig. \ref{modafshar}). The combined state
of the photon with its polarization, as it comes out of the two holes,
is given by
\begin{equation}
|\Psi_1\rangle = \tfrac{1}{\sqrt{2}}(|\psi_A'\rangle|L\rangle
+ |\psi_B'\rangle|R\rangle) ,
\end{equation}
where $|\psi_A'\rangle, |\psi_B'\rangle$ are the states corresponding to
the photon coming out of the hole $A$ and $B$, respectively, and 
$|L\rangle, |R\rangle$ are the left- and right-circular polarization states,
respectively. As the photon travels and reaches the lens, or the location
of a potential screen, the states $|\psi_A'\rangle, |\psi_B'\rangle$
broaden and overlap with each other. The state at this time can be written as
\begin{equation}
|\Psi_2\rangle = \tfrac{1}{\sqrt{2}}(|\psi_A\rangle|L\rangle
+ |\psi_B\rangle|R\rangle) ,
\label{psi2}
\end{equation}
where $|\psi_A\rangle, |\psi_B\rangle$ are the states corresponding to
the photon coming from the hole $A$ and $B$, respectively. This is an
entangled state, and has an important implication. If the photon is 
found in the polarization state $|L\rangle$ ($|R\rangle$), it implies
that it came from the hole $A$ ($B$). So the polarization of the photon
can be used any time, to determine the path it took, as long as this
entangled state is intact. In the absence
of the quarter wave plates, the state would simply have been
\begin{equation}
|\Psi_2'\rangle = \tfrac{1}{\sqrt{2}}(|\psi_A\rangle + |\psi_B\rangle) .
\label{psi2p}
\end{equation}

Now if the photons are to show interference, there should be some parts
of the amplitudes $|\psi_A\rangle, |\psi_B\rangle$ which cancel with each
other, to give destructive interference. This is an essential requirement
of an interference pattern. So we assume the two states to have the
following form
\begin{equation}
|\psi_A\rangle = \tfrac{1}{\sqrt{2}}( |\phi_+\rangle + |\phi_-\rangle ), \hskip 5mm
|\psi_B\rangle = \tfrac{1}{\sqrt{2}}( |\phi_+\rangle - |\phi_-\rangle ) ,
\label{phi}
\end{equation}
where we make no assumption on the form of $|\phi_+\rangle, |\phi_-\rangle$,
except that they are orthonormal. The part $|\phi_+\rangle$ will constitute
the bright fringes, and the part $|\phi_-\rangle$, rather its absence,
will constitute the dark fringes. Note that $|\psi_A\rangle$ and
$|\psi_B\rangle$ are orthogonal as they result from the same unitary
evolution from the orthogonal initial states $|\psi_A'\rangle,
|\psi_B'\rangle$.

One may wonder if introduction of thin wires would lead to photons scattering
off the wires. Indeed it will, and for that reason we assume that the wires
are fully absorptive, and do not reflect (scatter) photons. However, even
if the wires are fully absorptive, introduction of these would be like 
introducing a grating, and \emph{diffraction} effects would be there.
This point has been discussed in detail by Jacques et al. \cite{jacques}.
However, the experiments in question did acknowledge that if only one
hole is open, the photons will be scattered/absorbed, and that will reduce
the photon count at the detector. The central argument of these experiments
uses just the fact that when only one hole is open, only one detector receives 
photons, in the absence of any wires. In our analysis, we will not include
the effect of thin wires explicitly, but just use the fact that their presence
may affect the photon count, and that can be used to infer the presence or
absence of interference. Our path distinguishability analysis will be
in the absence of the wires, as the experiments in question claim that
in the absence of wires, with the both the holes open, the two detectors
give which-hole information.

Next we need to incorporate the effect of the lens on the photon. Without
going into the details of how a lens works, here it suffices to consider
its effect on $|\psi_A\rangle$ and $|\psi_B\rangle$. We know that the lens
takes the state $|\psi_A\rangle$ to the detector $D_A$ and $|\psi_B\rangle$
to the detector $D_B$. So, the effect of the lens can be treated as a unitary
operator with the effect
\begin{eqnarray}
U_L|\psi_A\rangle &=& \tfrac{1}{\sqrt{2}}U_L( |\phi_+\rangle + |\phi_-\rangle)
 = |D_A\rangle \nonumber\\
U_L|\psi_B\rangle &=& \tfrac{1}{\sqrt{2}}U_L( |\phi_+\rangle - |\phi_-\rangle) 
 = |D_B\rangle ,
\label{phip}
\end{eqnarray}
where $|D_A\rangle$ ($|D_B\rangle$) is the state of the photon if it registers
at the detector $D_A$ ($D_B$).
Now we wish to see the effect of the lens on the state (\ref{psi2}), which
has photon polarization included. The state of the photon when it arrives
at the two detectors, is given by
\begin{eqnarray}
|\Psi_3\rangle &=& U_L|\Psi_2\rangle
= \tfrac{1}{\sqrt{2}}U_L(|\psi_A\rangle|L\rangle
+ |\psi_B\rangle|R\rangle) \nonumber\\
 &=& \tfrac{1}{\sqrt{2}}(|D_A\rangle|L\rangle + |D_B\rangle|R\rangle) .
\label{psi3}
\end{eqnarray}
It is clear that the photons that arrive at $D_A$, have the polarization
$|L\rangle$, and those that arrive at $D_B$, have the polarization
$|R\rangle$, which means those that landed at $D_A$ came from hole $A$,
and those that arrived at $D_B$ came from hole $B$. However, these photons
will not show any interference. This can be inferred from the fact that
the photons that landed at $D_A$, with polarization $|L\rangle$,
had the state $\tfrac{1}{\sqrt{2}}( |\phi_+\rangle + |\phi_-\rangle)$
before the lens. This state has the part $|\phi_-\rangle$ present, which
means the photons pass through the regions which would be dark if there
were interference.

Now we look for situations which \emph{can} give us interference. 
The circular polarization states can also be written in terms of 
horizontal and vertical polarization states:
\begin{equation}
|R\rangle = \tfrac{1}{\sqrt{2}}(|H\rangle + i|V\rangle), \hskip 5mm
|L\rangle = \tfrac{1}{\sqrt{2}}(|H\rangle - i|V\rangle),
\label{HV}
\end{equation}
where $|H\rangle, |V\rangle$ are the horizontal and vertical polarization
states, respectively. Using (\ref{HV}), the state before the lens
(\ref{psi2}) can be written as
\begin{equation}
|\Psi_2\rangle = \tfrac{1}{{2}}(|\psi_A\rangle + |\psi_B\rangle)|H\rangle
- \tfrac{i}{{2}}(|\psi_A\rangle - |\psi_B\rangle|V\rangle) .
\label{psi2hv}
\end{equation}
This entangled state implies that if the polarization state of the photon
is found to be $|H\rangle$, it implies that it came from \emph{both the holes}.
The same holds true for photons found in the polarization state $|V\rangle$.
The photon state, before the lens, can also be written as 
\begin{equation}
|\Psi_2\rangle = \tfrac{1}{\sqrt{2}}(|\phi_+\rangle|H\rangle
- i|\phi_-\rangle|V\rangle) .
\label{psi2hv1}
\end{equation}
The photon state correlated to $|H\rangle$ polarization does not have the
$|\phi_-\rangle$ part, which means it will show interference. So the
conclusion is that photons which are found in the horizontal polarization
state, will show interference.
We can filter out such photons by putting a horizontal polarizer in front
of the detectors $D_A$ and $D_B$ (see Fig. \ref{modafshar}).
However, first we would like to see the effect of the lens on these
photons. From (\ref{phip}) we see that
\begin{equation}
U_L|\phi_+\rangle = \tfrac{1}{\sqrt{2}}(|D_A\rangle + |D_B\rangle ,
\end{equation}
which means that the lens will take the photons with horizontal polarization
to both the detectors, and not just one of the two. The final state of the
photon at the detectors will be
\begin{eqnarray}
|\Psi_3\rangle &=& U_L|\Psi_2\rangle
= \tfrac{1}{\sqrt{2}}U_L(|\phi_+\rangle|H\rangle
- i|\phi_-\rangle|V\rangle) \nonumber\\
 &=& \tfrac{1}{{2}}(|D_A\rangle + |D_B\rangle)|H\rangle 
 - i\tfrac{1}{{2}}(|D_A\rangle - |D_B\rangle)|V\rangle . \nonumber\\
\label{psi3p}
\end{eqnarray}
The above state implies that the photons with horizontal polarization are
equally likely to land at either of the two detectors.
The photons that pass through the horizontal polarizer, will have the state
\begin{equation}
\langle H|\Psi_3\rangle = \tfrac{1}{{2}}(|D_A\rangle + |D_B\rangle) ,
\end{equation}
meaning they can land at either of the two detectors. More importantly,
their polarization state is $|H\rangle$ which, by looking at (\ref{psi2hv})
implies that these photons, which show interference, \emph{came from both 
the holes}. Thus all the photons landing at (say) $D_A$, with horizontal
polarization, passed through both the holes. This goes against the conclusion of
the modified two-slit interference experiments that each detectors tells
us which hole the photon came from.

The reader might have guessed that an equivalent analysis can be done
for photons with vertical polarization. Such photons will also show 
interference, but the interference pattern will be shifted such that
the intensity minima will lie at the locations of the maxima of the
photons with horizontal polarization. So, if one wants to test out the
presence of interference in such a situation, the wires need to be inserted
at different locations. The reader might also have recognized this 
phenomenon as the familiar \emph{quantum erasure} \cite{eraser}.

\section{Conclusion}

We have analyzed a typical modified two-slit experiment by introducing quantum
path markers. The polarization state of the photon is used as the path marker.
The analysis shows that in this experiment, if the photons retain 
which-slit information, they do not show interference, but always land at
a particular detector. The photons which do show interference, may land
at any of the two detectors. However, their path marker shows that such
photons always come from both the slits, and not one of the two, even though
they eventually land at one of the two detectors at random. So for photons
which show interference, their landing at a particular detector does 
\emph{not} imply that they came from a particular slit. This shows that the
assumptions in the much debated modified two-slit experiments, that the
each detector gives information on which slit the photon came from, is
incorrect. The conclusion of the present work is arrived at, not by any
philosophical argument or assumptions, but by experimentally verifiable
correlation between the photon paths and path marker states. Our modified
experiment with path markers can be performed easily, since a quantum eraser
using photon polarization was demonstrated long back \cite{walborn}.

\section*{Acknowledgements}
The authors wishes to thank the two anonymous referees for their various
suggestions which led to much improved clarity and readability of the
paper. The author is thankful to  Alexandra Elbakyan for her support.

\section*{Declaration}

The author has no conflict of interest.


\begin{thebibliography}{0}

\bibitem{bohr} N. Bohr, The quantum postulate and the recent development of atomic theory.
\href{ https://doi.org/10.1038/121580a0}
{Nature (London)  {121}, 580 (1928).}

\bibitem{feynman} R.P. Feynman, R.B. Leighton, M. Sands,
Lectures on Physics, Vol. 3, pp. 1-1 (Addison-Wesley, 1966)

\bibitem{tqrv} T. Qureshi, R. Vathsan, Einstein's recoiling slit experiment, complementarity and uncertainty.
\href{https://doi.org/10.12743/quanta.v2i1.11}
{Quanta 2, 58 (2013).}

\bibitem{triality1} T. Qureshi, Predictability, distinguishability, and entanglement.
\href{https://doi.org/10.1364/OL.415556}
{\emph{Opt. Lett.} {46}, 492 (2021).}

\bibitem{triality2} A.K. Roy, N. Pathania, N.K. Chandra, P.K. Panigrahi, T. Qureshi, Coherence, path predictability, and I concurrence: A triality.
\href{https://doi.org/10.1103/PhysRevA.105.032209}
{Phys. Rev. A 105, 032209 (2022).}

\bibitem{afshar1} S.S. Afshar, E. Flores, K.F. McDonald, E. Knoesel,
Paradox in wave-particle duality.
\href{https://doi.org/10.1007/s10701-006-9102-8}
{Found. Phys. 37, 295 (2007).}

\bibitem{afshar2} E.V. Flores, Modified Afshar experiment: calculations. In: Roychoudhuri, C., Kracklauer, A. F., Khrennikov, A.Y. (eds)
\href{https://doi.org/10.1117/12.826015}
{The Nature of Light: What are Photons? III Proc. SPIE, vol. 7421. San Diego, SPIE pp. 74210W (2009).}

\bibitem{chown} M. Chown, ``Quantum rebel,"
{\emph{New Scientist} {183}, 30 (2004).}

\bibitem{kwait} P.G. Kwiat, H. Weinfurter, T. Herzog, A. Zeilinger, and
M.A. Kasevich, Interaction-free measurement.
\href{https://doi.org/10.1103/PhysRevLett.74.4763}
{Phys. Rev. Lett.  74, 4763 (1995).}

\bibitem{kastner} R. E. Kastner, Why the Afshar Experiment Does Not
Refute Complementarity.
\href{https://doi.org/10.1016/j.shpsb.2005.04.006}
{Studies In History and Philosophy of Science Part B 36 6498 (2005).}

\bibitem{srinivasan} R. Srinivasan, Logical analysis of the Bohr
Complementarity Principle in Afshar's experiment under the NAFL
interpretation. 
\href{https://doi.org/10.1142/S021974991000640X}
{Int. J. Quant. Inf. 8, 465 (2010).}

\bibitem{drezet} A. Drezet, Wave particle duality and Afshar's experiment.
{Progress Phys., 7(1), 57 (2011).}

\bibitem{steuernagel} O. Steuernagel, Afshar's Experiment does not show a
Violation of Complementarity.
\href{https://doi.org/10.1007/s10701-007-9153-5}
{Found. Phys. 37, 1370 (2007).}

\bibitem{georgiev} D.D. Georgiev, Single photon experiments and quantum complementarity.
{Progress Phys. 2, 97–103 (2007).}

\bibitem{flores} E.V. Flores, Reply to Comments of Steuernagel on the Afshar's Experiment.
\href{https://doi.org/10.1007/s10701-008-9234-0}
{Found. Phys. 38, 778 (2008).}

\bibitem{jacques} V. Jacques, N. D. Lai, A. Dréau, D. Zheng, D. Chauvat, F. Treussart, P. Grangier, and J.-F. Roch, Illustration of quantum complementarity using single photons interfering on a grating.
\href{https://doi.org/10.1088/1367-2630/10/12/123009}
{New J. Phys. 10, 123009 (2008).}

\bibitem{tq} T. Qureshi, Modified two-slit experiments and complementarity.
\href{https://doi.org/10.4236/jqis.2012.22007}
{J. Quantum Inf. Sci. 2, 34 (2012).}

\bibitem{tata} E.V. Flores, J.M. De Tata, Complementarity Paradox Solved: Surprising Consequences.
\href{https://doi.org/10.1007/s10701-010-9477-4}
{Found. Phys. 40, 1731 (2010).}

\bibitem{kaloyerou} P. N. Kaloyerou, Critique of quantum optical experimental refutations of Bohr’s principle of complementarity, of the Wootters–Zurek principle of complementarity, and of the particle–wave duality relation.
\href{https://doi.org/10.1007/s10701-015-9959-5}
{Found Phys 46, 138 (2015).}

\bibitem{knight} A. Knight, No paradox in wave–particle duality.
\href{https://doi.org/10.1007/s10701-020-00379-9}
{Found. Phys. 50, 1723 (2020).}

\bibitem{batelaan} B. Gergely, H. Batelaan, Simulation of Afshar's double slit experiment.
\href{https://doi.org/10.1007/s10701-022-00585-7}
{Found. Phys. 52, 69 (2022).}

\bibitem{eraser} M. O. Scully and K. Dr\"{u}hl, Quantum eraser: A proposed photon correlation experiment concerning observation and ``delayed choice" in quantum mechanics.
\href{https://doi.org/10.1103/PhysRevA.25.2208}
{Phys. Rev. A 25, 2208 (1982).}

\bibitem{walborn} S. P. Walborn, M. O. Terra Cunha, S. Pádua, C. H. Monken,
``Double-slit quantum eraser,"
\href{https://doi.org/10.1103/PhysRevA.65.033818}
{Phys. Rev. A 65, 033818 (2002).}


\end{thebibliography}
\end{document}